\documentstyle[aps,epsf]{revtex}
 
\newcommand{\bq}{\begin{equation}}
\newcommand{\eq}{\end{equation}}
\newcommand{\bqn}{\begin{eqnarray}}
\newcommand{\eqn}{\end{eqnarray}}
\newcommand{\nb}{\nonumber}
\newcommand{\lb}{\label}

\begin{document}

\title{ On the Levi-Civita solutions with cosmological constant} 
\author{ M.F.A. da Silva, Anzhong Wang, and Filipe M. Paiva}
\address{Departamento de F\' {\i}sica Te\' orica,
Universidade do Estado do Rio de Janeiro,
Rua S\~ ao Francisco Xavier, 524, Maracan\~a, CEP. 20550-013, 
Rio de Janeiro, RJ, Brazil}
\author{N.O. Santos}
\address{Centro Regional Sul de Pesquisas Espaciais - INPE/MCT,
Laborat\'orio de Astrof\'{\i}sica e Radioastronomia,
Universidade Federal de Santa Maria - LACESM,
Cidade Universit\'aria, CEP. 97105-900, Santa Maria, RS, Brazil}

\date{ November 3, 1999}

\maketitle

\begin{abstract}

The main properties of the Levi-Civita solutions with the cosmological constant
are studied. In particular, it is found that some of the
solutions need to be extended beyond certain hypersurfaces in order to have
geodesically complete spacetimes. Some extensions  are considered and found 
to give rise to black hole structure but with plane symmetry.
All the spacetimes that are not geodesically complete are Petrov type $D$,
while in general the spacetimes are Petrov type $I$. 
 
\end{abstract}

\vspace{.6cm}

PACS numbers: 04.20Jb; 04.40.+c; 97.60.Lf.

\section{Introduction} 
 
Recently, we studied the Levi-Civita (LC) solutions and found that the
solutions have physical meaning at least for $ \sigma \in [0, 1]$,
where $\sigma $ is a free parameter related to the
mass per unit length   \cite{WSS1997}. 
In this paper, we shall  study the Levi-Civita solutions with the
cosmological  constant (LCC).  This is not trivial, as the inclusion of the
cosmological constant usually makes the problem considerably complicated and
changes  the spacetime properties dramatically. 

The paper is organized as follows:  In Sec. $II$ we shall study the main
properties of the LCC solutions, including their singularity behavior. 
We shall show that some
spacetimes are not geodesically complete and need to be extended.  In Sec.
$III$ we  will present some extensions and  show
that some of the extended spacetimes have  black hole  structure but with
plane symmetry. To distinguish these black holes with the spherical
ones, we shall refer them as {\em black membranes}.  
To further study the LCC
solutions, we devote Sec. $IV$ to investigate   their Petrov classifications,
while  Sec. $V$ contains our main  conclusions.

\section{The Main Properties of the Levi-Civita solutions with cosmological
constant}

\setcounter{equation}{0}
 
The LCC solutions are not new and were re-derived several times,
for example, see  \cite{Linet1986} and references therein. It can be
shown that, in addition to the cosmological constant, the solutions 
have {\em only} two physically relevant parameters, similar to the LC
solutions, and that, without loss of generality, they can be 
written  in the form,  
\bqn
\label{4.6}
ds^2 = Q(r)^{2/ 3}& &\left\{ P(r)^{-2(4\sigma^{2}-8\sigma +1)/ 3A}
{dt}^ 2 - P(r)^{2(8\sigma^{2}-4\sigma -1)/ 3A}dz^2 \right.\nonumber\\
&&\left. - C^{-2} P(r)^{-4(2\sigma^{2}+2\sigma -1)
/ 3A}d\varphi^2\right\} - dr^2,
\eqn
where $\{x^{\mu}\} \equiv \{t, r, z, \varphi\}$
are the usual cylindrical  coordinates, $A \equiv 4\sigma ^{2} -
2\sigma + 1$. The constant $\sigma$ is related, but not equal, 
to the mass  per
unit length, and $C$ is related to the angle defects 
\cite{WSS1997,Bonnor1992}. The
functions  $P(r)$ and $Q(r)$ are defined as, 
\bq
\lb{4.7}
P(r) \equiv \frac{2}{\sqrt{3\Lambda}}
\tan\left(\frac{\sqrt{3\Lambda} r}{2}\right),\;\;\;
Q(r) \equiv \frac{1}{\sqrt{3\Lambda}}
\sin\left(\sqrt{3\Lambda} r\right).
\eq
 It is easy to show that as 
$\Lambda \rightarrow 0 $ the above solutions 
reduce to the LC solutions \cite{Kramer1980}. 
To study these solutions, it is found convenient to consider the two
cases, $\Lambda > 0$ and $\Lambda < 0$, separately.

\subsection{$\Lambda > 0$}

In this case, from Eq.(\ref{4.7}) we find that, as $r \rightarrow 0$,
we have 
$
Q(r) \approx r, P(r) \approx r.
$
Then, the corresponding solutions approach to 
the LC ones. As a result, the metric (\ref{4.6}) has the same singularity
behavior as the LC ones near the axis $r = 0$. In particular, for the
cases $\sigma = 0$ and $\sigma = 1/2$, the solutions are
free of spacetime singularities \cite{WSS1997}. Thus, one may consider
the LCC solutions with $\sigma = 0,\; 1/2$ as cylindrical analogues
of the de Sitter solution, although there is a foundamental difference
between these two cases. In the present case 
the Weyl tensor is different from zero, and
the spacetimes are not conformally flat [cf. the discussions in Sec. IV].
As a matter of fact, they are all Petrov type $D$. In addition
to the usual three Killing vectors, $\xi_{(t)} = \partial t, \;
\xi_{(z)} = \partial z,\; \xi_{(\varphi)} = \partial \varphi$,   
the solution
with $\sigma = 0$ has one more Killing vector, $\xi_{(0)} = t\partial z -
z\partial t$, which  corresponds to a Lorentz boost in the $tz$-plane, while
the solution  with $\sigma = 1/2$ has the fourth Killing vector given by
$\xi_{(1/2)} =  C^{-1}\varphi\partial z -  C z\partial \varphi$, which
corresponds  to a rotation in the $\varphi z$-plane. Since in the latter case
the extrinsic  curvatures of the two spacelike surfaces $r = Const.$ and $t =
Const.$ are identically zero, it is difficult to consider  this spacetime as
having cylindrical symmetry. Instead, one may extend the $\varphi$-coordinate
from the range [$0, 2\pi$] to the range ($-\infty, + \infty$), so the
resulting spacetime has  plane symmetry \cite{WSS1997}.  

On the other hand, Eq.(\ref{4.7}) shows that the solutions usually
are also singular on the hypersurface 
$
r = r_{g} \equiv \pi/\alpha,
$
where $\alpha \equiv (3|\Lambda|)^{1/2}$. To study the singular behavior of
the solutions near this hypersurface, we use the relations
$
Q(r) \approx R,  P(r) \approx {R}^{-1},
$
as $r \rightarrow r_{g}$, where $ R \equiv r - r_{g}$. 
Substituting these expressions 
into Eq.(\ref{4.6}), we find  
\bq
\label{4.13}
ds^2 \approx  R^{4(4\sigma^{2} - 5\sigma +1)/3A}  
{dt}^ 2
- R^{-4(2\sigma^{2}-\sigma -1)/ 3A}dz^2 
- C^{-2} R^{2(8\sigma^{2}+2\sigma -1)
/ 3A}d\varphi^2 - dr^2, \; ( R \approx 0).  
\eq
The corresponding
Kretschmann scalar is given by  
\bq
\lb{4.14}
 R^{\alpha \beta \gamma \delta}
R_{\alpha \beta \gamma \delta} =  \frac {64 
(\sigma  - 1)^{2}(2\sigma  + 1)^{2
}(4\sigma  - 1)^{2}}{27A^{3}R^{4}}, \; ( R \approx 0),
\eq
which is always singular as $R \rightarrow 0$, except for the cases 
where $\sigma = -1/2, \; 1/4,\;
1$.  It can be shown that all the fourteen scalars built from the
Riemann tensor have the same properties. Therefore, in the cases
$\sigma = -1/2, \; 1/4,\; 1$ the singularities on the hypersurface
$r = r_{g}$ are coordinate ones, and to have the corresponding
spacetimes geodesically complete, the solutions need to be extended
beyond this surface.  
Note that, similar to the solutions with $\sigma = 0,\; 1/2$,  
all these three solutions  are 
Petrov type $D$. Moreover, in addition to the usually three Killing
vectors, they also have the fourth Killing vector, given, respectively, by 
$\xi_{(-1/2)} = C^{-1}\varphi\partial z -  C z\partial \varphi$,  
$\xi_{(1/4)} = C^{-1}\varphi\partial t -  C t\partial \varphi$,  and 
$\xi_{(1)} = z \partial t -  t \partial z$. Using the same arguments
as those given for the solution with $\sigma = 1/2$, the solution  
with $\sigma = -1/2$ can be also considered as representing plane symmetry. 

Combining the analysis of the singular behavior of the solutions 
on the axis and on the hypersurface $r = r_{g}$, we can see that 
all the solutions are singular on both of the two surfaces, except
for the ones with $\sigma = -1/2, \; 0, 1/4,\; 1/2,\; 1$, {\em which
are the only solutions that are Petrov type D} [See the discussions given 
in Sec. IV]. These
singularities make the physical interpretation of the solutions
very difficult. A possible way to circuit these difficulties
is to cut the spacetimes along the hypersurface $ r = r_{\Sigma} 
< r_{g}$, and then join the part $ r < r_{\Sigma}$ with an asymptotically
de Sitter region, while considering the singularities on the axis as
representing matter sources \cite{Bonnor1992}. On the other hand, 
the solutions
with $\sigma = 0,\; 1/2$ are free of spacetime singularities
on the axis, but do have on the hypersurface $r = r_{g}$. To
give a meaningful physical interpretation of these solutions,
one may take $r = r_{g}$ as the symmetry axis, and then extend
the spacetimes beyond $r = 0$. When $\sigma = -1/2,\; 1/4,\;
1$, the corresponding solutions are singular on the axis, but
free of spacetime singularities on the hypersurface $r = r_{g}$.
Thus,  we need to
extend the spacetimes beyond this surface.  We shall leave
these considerations to the next section.

\subsection{$\Lambda < 0$}

 As $r \rightarrow 0$ the functions $Q(r)$ and $P(r)$ have
the same asymptotical behavior as those given in the last case. As
a result, in both of the two cases the solutions have the same singularity
behavior as the LC ones near the axis $r = 0$, that is, they are 
all singular, except for the cases $\sigma = 0$ and $\sigma = 1/2$.
  
On the other hand, Eq.(\ref{4.7}) shows that in the present case
$Q(r)$ and $P(r)$ are monotonically increasing functions of $r$ 
and are positive for any given $r > 0$, in contrast to the case 
$\Lambda >
0$, where they are periodic functions [cf. Eq.(\ref{4.7})]. 
When $r \rightarrow + \infty$, we find
$
Q(r) \approx e^{\alpha r}/({2 \alpha}),\;
P(r) \approx {2}/{ \alpha},
$
and the corresponding metric, after $t$ and $z$ are rescaled,
takes the form
\bq
\label{5.4}
ds^2 \approx  C_{0} e^{2\alpha r/3} \left(  dt^{2}
- dz^{2} - C^{-2}d\varphi^{2}\right) - dr^{2},\;\;
(r \rightarrow + \infty),
\eq
where $C_{0}$ is a positive constant. This is
exactly the anti-de Sitter spacetime but written in the horo-spherical
coordinates \cite{CGS1993}. Since the metric does not depend on 
the parameter $\sigma$,
we conclude that {\em all the LCC solutions with negative cosmological
constant are asymptotically anti-de Sitter}.

\section{Solutions Representing Black Membranes}

As shown in the last section, the solutions with $\sigma = \pm 1/2$
both for $\Lambda > 0$ and $\Lambda < 0$ have the fourth Killing
vector, $\xi = C^{-1}\varphi\partial z -  Cz\partial \varphi$,
which represents the rotation invariant in the $z\varphi$-planes, or
in other words, the extrinsic curvature of the planes is identically
zero. This property makes these two dimensional planes more
like of having plane symmetry than  cylindrical one \cite{WSS1997}. Then,
the ranges of $r$ and $\varphi$ should be extended to
$- \infty < r, \; \varphi < + \infty$. In the following we shall
denote such extended coordinate $\varphi$ by $Y$. Once this is done, 
we can see that the spacetime is not geodesically complete. 
In particular, the solutions with $\sigma = 1/2$ for both
$\Lambda > 0$ and $\Lambda < 0$ are not singular
on the hypersurface $r = 0$ and need to be extended beyond
it, while the one with $\sigma = -1/2$ and $\Lambda > 0$ is not singular
on the hypersurface $r = r_{g}$ and needs to be extended beyond this
surface, too. In the following, we shall consider these cases separately.

{\bf Case $\alpha$) $\; \sigma = 1/2,\; \Lambda > 0$}:
In this case,  making  the  coordinate transformations
\bq
\lb{6.1}
T = \frac{2t}{3},\;\; 
X = \cos^{2/3}\left(\frac{\alpha r}{2}\right), \;\; 
Y = \frac{\alpha \varphi}{3 C},\;\; Z = \frac{\alpha z}{3},
\eq
we find that the corresponding solution  can be written in the form,  
 \bq
\lb{6.2}
ds^2_{\sigma = 1/2}  
 =\frac{9}{\alpha^{2}}\left\{
        f(X) d{T}^{2} - f^{-1}(X) dX^{2} 
        - X^{2}\left(dY^{2} + d{Z}^{2}\right)\right\},
        \;(\Lambda > 0),
\eq
where $f(X)$ is defined as
\bq
\lb{6.3}
f(X) \equiv \frac{1}{X} - X^{2}.
\eq
From Eq.(\ref{6.1}) we can see that the region $ 0 \le r \le r_{g}$
is mapped into the region $  0 \le X \le 1$, and the point $ r= r_{g}$,
 where the spacetime is singular, is mapped to the point $X = 0$.
Extending $X$ to the range ($-\infty, \; +\infty$), we find that
in the extended spacetime two new regions, $X > 1$
and $X < 0$, are included. The curvature singularity at $X = 0$
divides the whole spacetime into two unconnected regions, $X \ge 0$ and
$X \le 0$. In the region $X \le 0$, the function $f(X)$ is always negative,
and the $X$ coordinate is timelike. Then, the spacetime is essentially 
time-dependent, and the singularity at $X =0$ is spacelike and naked. As $ X
\rightarrow - \infty$, the metric is asymptotically de Sitter \cite{HE1973},
\bq
\lb{6.4}
ds^2_{\sigma = 1/2}  
 \approx   d{\tilde{T}}^{2} - e^{2\alpha\tilde{T}/3}\left(dX^{2} 
        + dY^{2} + d{Z}^{2}\right),
        \;(X \rightarrow - \infty),
\eq
where $T = e^{\alpha\tilde{T}/3}$ and $X,\; Y,\; Z$ have
been rescaled. The corresponding Penrose diagram
is given by Fig.1(a).

When $X \ge 0$, $f(X)$ is greater than zero for $ 0 \le X < 1$ and less
than zero for $X > 1$, that is, $X$ is spacelike when $ 0 \le X < 1$ and 
timelike when  $X > 1$. On the hypersurface $X = 0$ it becomes null,
which represents a horizon. Since the spacetime singularity at $X = 0$
now is timelike, the horizon is actually a Cauchy horizon. As
$X \rightarrow + \infty$, the spacetime is also asymptotically de Sitter
and approaches the same form as that given by Eq.(\ref{6.4}).
The corresponding Penrose diagram is given by Fig.1(b). 

{\bf Case $\beta$)$\; \sigma = -1/2,\; \Lambda > 0$}: In this case, 
the spacetime is singular at $ r= 0$ and is free of curvature singularity
at $r = r_{g}$. Thus, to have a geodesically complete spacetime, we need to
extend the solution beyond the hypersurface $r = r_{g}$. To make such
an extension, we can introduce a new coordinate, $X$, by $
X = \sin^{2/3}(\alpha r/2)$ and  rescale the coordinates, $t, z$
and $\varphi$, then we will find that the corresponding metric  takes the
same form as that given by Eq.(\ref{6.2}). 
This is not expected. As
we know, in the limit $\Lambda \rightarrow 0$ the solution with $\sigma
= 1/2$ approaches the Rindler space \cite{Rindler1977}, which represents 
a uniformly gravitational field and is free of any kind of 
spacetime curvature singularities,  while the one with $\sigma = -1/2$
is the static Taub solution with plane symmetry \cite{Taub1951}, and
is singular on the hypersurface $X = 0$. The total mass of the
Taub spacetime is negative, while the one of Rindler is not
\cite{SWS1998}. However, the presence of the cosmological constant
makes up these differences and turns the two spacetimes identical!

{\bf Case $\gamma$)$\; \sigma = 1/2,\; \Lambda < 0$}: In this case, 
the spacetime is free of curvature singularity
for $0 \le r < + \infty$, and needs to be extended 
beyond the hypersurface $r = 0$. Similar to the last two cases, 
introducing the
new coordinate,  $X$, as $ X = \cosh^{2/3}(\alpha r/2)$,
and rescaling the cordinates $t, z, \varphi$,  
 the corresponding metric can be wiritten in the form,  
\bq
\lb{6.7}
ds^2_{\sigma = 1/2}  
 = \frac{9}{\alpha^{2}}\left\{ -
        f(X) d{T}^{2} + f^{-1}(X) dX^{2} 
        - X^{2}\left(dY^{2} + d{Z}^{2}\right)\right\},
        \;(\Lambda < 0),
\eq
where $f(X)$ is given by Eq.(\ref{6.3}). From the expression of $X$ we can 
see that the region $0 \le r < +\infty$ is mapped into the region
$1 \le X < + \infty$. The region $X < 1$ is an extended region.
After the extension, a spacetime curvature singularity appears at
$X = 0$, which divides the whole $X$-axis into two parts, $X \le 0$
and $X \ge 0$. It can be shown that, unlike the case $\Lambda > 0$, 
now the spacetime is static in the region 
$X \le 0$, and the curvature singularity at $X = 0$ is timelike 
and naked. As $X \rightarrow - \infty$, the spacetime is asymptotically
anti-de Sitter \cite{CGS1993},
\bq
\lb{6.7a}
ds^2_{\sigma = 1/2}  
 \approx   \frac{9}{\alpha^{2}\tilde{X}^{2}}\left(d{{T}}^{2} 
        -  d\tilde{X}^{2} 
        - dY^{2} - d{Z}^{2}\right),
        \;(X \rightarrow - \infty),
\eq
where $\tilde{X} = 1/X$. The corresponding Penrose diagram is 
given by Fig.2(a).

In the region $X \ge 0$, the spacetime singularity at $X = 0$ becomes
spacelike. Except for this curvature singularity, there is a coordinate one
located at $X = 1$. This coordinate singularity
actually represents an event horizon. As shown in the last section, 
the spacetime is asymptotically anti-de Sitter ($X \rightarrow +\infty$). 
The corresponding Penrose diagram is given by Fig.2(b). This is 
the black hole solution with plane symmetry found recently by Cai and 
Zhang with vanishing electromagnetic charge \cite{CZ1996}.

{\bf Case $\delta$)$\; \sigma = -1/2,\; \Lambda < 0$}: In this case, a
spacetime singularity appears at $r = 0$, and the whole region $ 0 \le r 
< + \infty$ is geodesically complete. However, 
since in this case the solution
has also  plane symmetry, and the range of $r$ should be taken as,
$- \infty < r < + \infty$. Then one may ask: what is the physical
interpretation of the  spacetime 
in the region $r \le 0$? To answer this question, let us introduce
a new coordinate  $
X = - \sinh^{2/3}(\alpha r/2)$, and rescale the other three, 
then we will find that
 the metric  takes the same form
as that  given by Eq.(\ref{6.7}). From the expression for
$X$ we can see that the 
region $0 \le r < +\infty$ now is mapped into the region $ - \infty <
X \le 0$,
while the region $- \infty < r \le 0$ is mapped to the region 
$0 \le X < +\infty$. In the region $0 \le r < +\infty$ the solution
represents a static spacetime with a naked singularity located at
$r = 0$. The spacetime is asymptotically anti-de Sitter, and the 
corresponding Penrose diagram is given by Fig.2(a). In the region 
$- \infty < r \le 0$ the solutions represents a black hole solution
with plane symmetry, and the corresponding Penrose diagram is given 
by Fig.2(b).

\section{The Petrov Classification of the Solutions}

To further study the LCC solutions, we shall
consider their Petrov classifications in this section.
Choosing a null tetrad, $e^{\mu}_{(\alpha)} = \{l^{\mu},\;
n^{\mu},\; m^{\mu},\;\bar{m}^{\mu}\}$, as
\bqn
\lb{7.1}
l^{\mu} &=& \frac{1}{\sqrt{2}}\left\{(g_{tt})^{1/2}\delta^{\mu}_{t}
+ \delta^{\mu}_{r}\right\},\nb\\
n^{\mu} &=& \frac{1}{\sqrt{2}}\left\{(g_{tt})^{1/2}\delta^{\mu}_{t}
- \delta^{\mu}_{r}\right\},\nb\\
m^{\mu} &=& \frac{1}{\sqrt{2}}\left\{(-g_{zz})^{1/2}\delta^{\mu}_{z}
+ i (-g_{\varphi\varphi})^{1/2}\delta^{\mu}_{\varphi}\right\},\nb\\
\bar{m}^{\mu} &=& \frac{1}{\sqrt{2}}\left\{(-g_{zz})^{1/2}\delta^{\mu}_{z}
- i (-g_{\varphi\varphi})^{1/2}\delta^{\mu}_{\varphi}\right\},
\eqn
where the metric coefficients can be read off directly 
from  Eq.(\ref{4.6}), 
we find that the non-vanishing components of the Ricci and Weyl tensors
are given by
\bqn
\lb{7.2}
R & = & 4\Lambda,\nb\\
\Psi_{0}&\equiv&-C_{\mu\nu\lambda\delta}l^{\mu}m^{\nu}
            l^{\lambda}m^{\delta}
= - \frac {\Lambda(4\sigma -1)}{4D^{2}\cos^{2}\theta\sin^{2}\theta}
\left[D\cos^{2}\theta + 2\sigma^{2} - \sigma - 1\right],\nb\\
\Psi_{2}&\equiv&-\frac{1}{2}C_{\mu\nu\lambda\delta}l^{\mu}m^{\nu}
           \bar m^{\lambda}n^{\delta}\nb\\
           &=& 
- \frac {\Lambda}{12D^{2}\cos^{2}\theta\sin^{2}\theta}
\left[D(8\sigma ^{2} - 4\sigma  - 1)\cos^{2}\theta 
- 32\sigma ^{3}(\sigma - 1) + 6\sigma ^{2} - 7\sigma   + 1\right], \nb\\
\Psi_{4}&\equiv&-C_{\mu\nu\lambda\delta}l^{\mu}m^{\nu}l^{\lambda}m^{\delta}
= \Psi_{0},
\eqn
where 
$
\theta \equiv {\sqrt{3\Lambda} r}/{2}.
$
Note that the above expressions are valid for any
 $\Lambda$, including $\Lambda = 0$.
  When $\Lambda < 0$ the function $\theta$
becomes imaginary, and the trigonometric functions become hyperbolic 
functions.  
Since $\Psi_0,\; \Psi_2$ and $\Psi_4$ are the only components of the
Weyl tensor different from zero, it can be shown that  
the metric in general is Petrov type $I$  \cite{Kramer1980}, 
unless i) $\Psi_0 = 0,\;
\Psi_{2} \not= 0$; ii)
$\Psi_0 = \pm 3\Psi_2 \not= 0$. In the last two 
cases, the solutions are
 Petrov type $D$. Further specialization $\Psi_0=\Psi_4 =
\Psi_2= 0$ leads to Petrov type  $O$ solutions. However,
the last case holds only when $\Lambda = 0$ and $\sigma = 0, 1/2$.
That is, all the solutions with $\Lambda \not= 0$ are either
Petrov type $I$ or $D$. From Eq.(\ref{7.2}) we find that the condition 
$\Psi_0 = 0$ and $\Psi_{2} \not= 0$ yields $\sigma = 1/4$, while the one
$\Psi_0 = \pm 3\Psi_2 \not= 0$ yields $\sigma = -1/2,\; 0,\; 1/2,\; 1$.
Thus, {\em all the solutions with $\Lambda \not= 0$
are Petrov type $I$, except for the ones with $\sigma = -1/2,\; 
0,\; 1/4,\; 1/2,\; 1$, which are Petrov type $D$}. In the latter
cases, all of the solutions have an additional Killing
vector [cf. Sec.II]. Since conformally flat solutions are necessarily
Petrov type $O$, we conclude that all the 
solutions with $\Lambda \not= 0$ are not conformally flat, and
the de Sitter and anti-de Sitter solutions are not particular
cases of the LCC solutions.  
 
It is interesting to note that if we introduce a new parameter $\tau$
by $\sigma = 1/4 + \tau$, we find that the metric can be obtained from the
one with $\sigma = 1/4 - \tau$ following the change,
$
t = i C^{-1}\varphi',\;\;
\varphi = i Ct'.
$
This indicates some kind of symmetries with respect to the solution
$\sigma = 1/4$. The study of the Ricci  and Weyl tensors  using
the null tetrad defined by Eq.(\ref{7.1}) 
will make this symmetry clear. For any given
$\tau$, we find   
\bqn
\lb{7.8}
R^{+}(r, \tau) &=& R^{-}(r, \tau),\;\;
\Psi^{+}_{0}(r, \tau) = - \Psi^{-}_{0}(r, \tau),\nb\\
\Psi^{+}_{2}(r, \tau) &=& \Psi^{-}_{2}(r, \tau),\;\;
\Psi^{+}_{4}(r, \tau) = - \Psi^{-}_{4}(r, \tau),
\eqn
where quantities with ``$+$" denote the ones calculated from the metric
with $\sigma = 1/4 + \tau$ and the quantities with ``$-$" denote
the ones calculated from the
metric with $\sigma = 1/4 - \tau$. The above relations
are valid even for $\Lambda = 0$. From Eq.(\ref{7.8}) we can 
see that, for any given $\tau$ the solution with $\sigma = 1/4 + \tau$ 
and the one with $\sigma = 1/4 - \tau$ have the same Petrov classification.
For example, the solution with $\sigma = 0$ and the one with 
$\sigma = 1/2$ all belong to Petrov type $D$ when $\Lambda \not= 0$, and
to Petrov type $O$ when $\Lambda = 0$.

\section{Conclusions} 

In this paper, we have studied the main properties of the Levi-Civita
solutions with the cosmological constant, and found that, among other things,
some solutions need to be extended beyond certain hypersurfaces in order
to obtain geodesically complete spacetimes. We have considered some extensions
for the case where the solutions have a rotating Killing vector in the
$z\varphi$-plane, and found that some of the extensions give rise to black
hole structures but with plane symmetry, {\em black membranes}. 
It is interesting to note that these structures  
exist even in the range, $ - \infty < r
\le 0$. This naturally raises the question: What kind of spacetimes do the
general solutions represent in this region? This problem is currently
under our  investigation.

To further study the solutions, we have also considered their Petrov
classifications, and found that all the solutions that are not 
geodesically complete, including the ones that represent black  
membranes, are Petrov type $D$, while in general they are
Petrov type $I$. As we know, the Kerr-Newmann
solutions are Petrov type $D$, too. So, it would be 
very interesting to show that all the black hole solutions
with plane or cylindrical symmetry  
are Petrov type $D$.

\section*{Acknowledgment} 

FMP thanks Jim Skea for useful discussions. 
The financial assistance from CNPq (AW, NOS),
and the one from FAPERJ (AW, FMP, MFAS) are gratefully 
acknowledged. 



\vspace{2.cm}

\section*{Figure captions}

Fig.1  The Penrose diagram for the cases $\sigma = \pm 1/2,\;
\Lambda > 0$. (a) $X \le 0$; (b) $X \ge 0$. In the figure, each point
actually represents a plane. The lines
$X = 0$ represent spacetime singularities, 
while the lines $X = 1$ represent Cauchy horizons. As $|X| \rightarrow
+ \infty$, the spacetimes are asymptotically de Sitter.

Fig.2  The Penrose diagram for the cases $\sigma = \pm 1/2,\;
\Lambda < 0$. (a) $X \le 0$; (b) $X \ge 0$.  The lines
$X = 0$ represent spacetime singularities, 
while the lines $X = 1$ represent event horizons.
As $|X| \rightarrow
+ \infty$, the spacetimes are asymptotically anti-de Sitter.

\newpage

\epsfysize=20cm

\centerline{\epsfbox{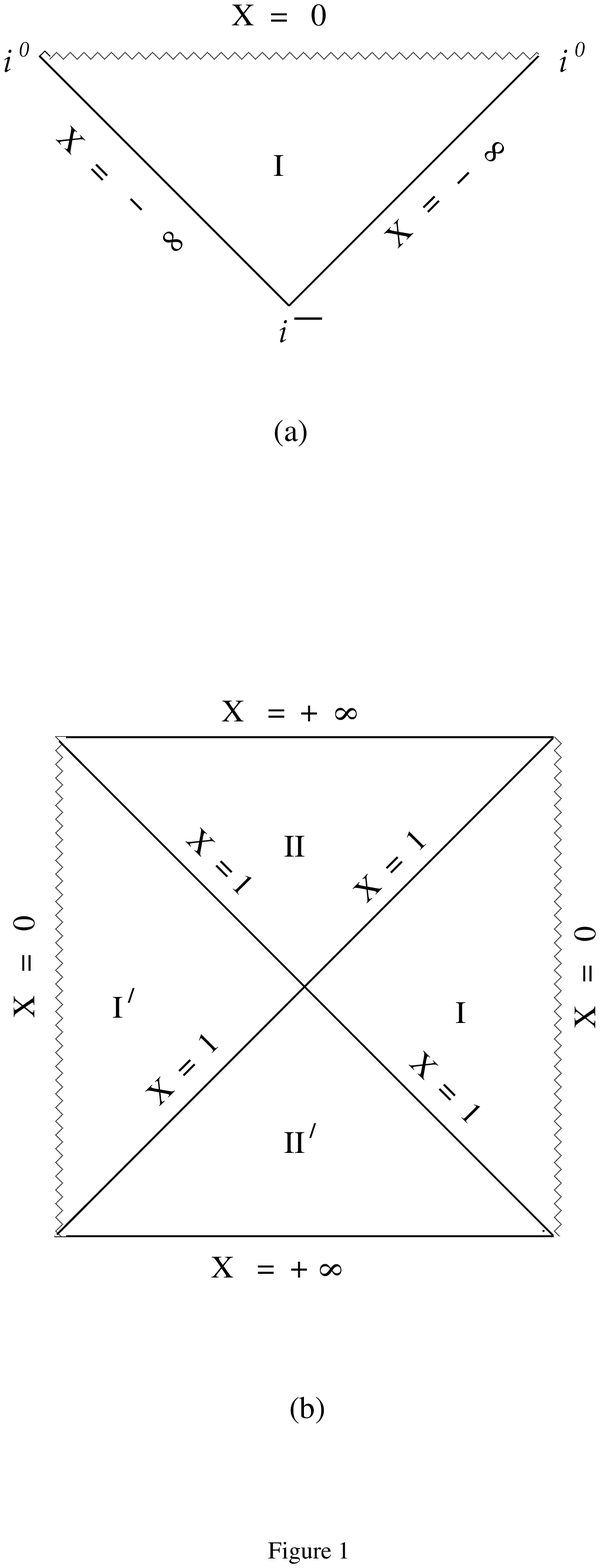}}

\newpage
\epsfysize=23cm
\centerline{\epsfbox{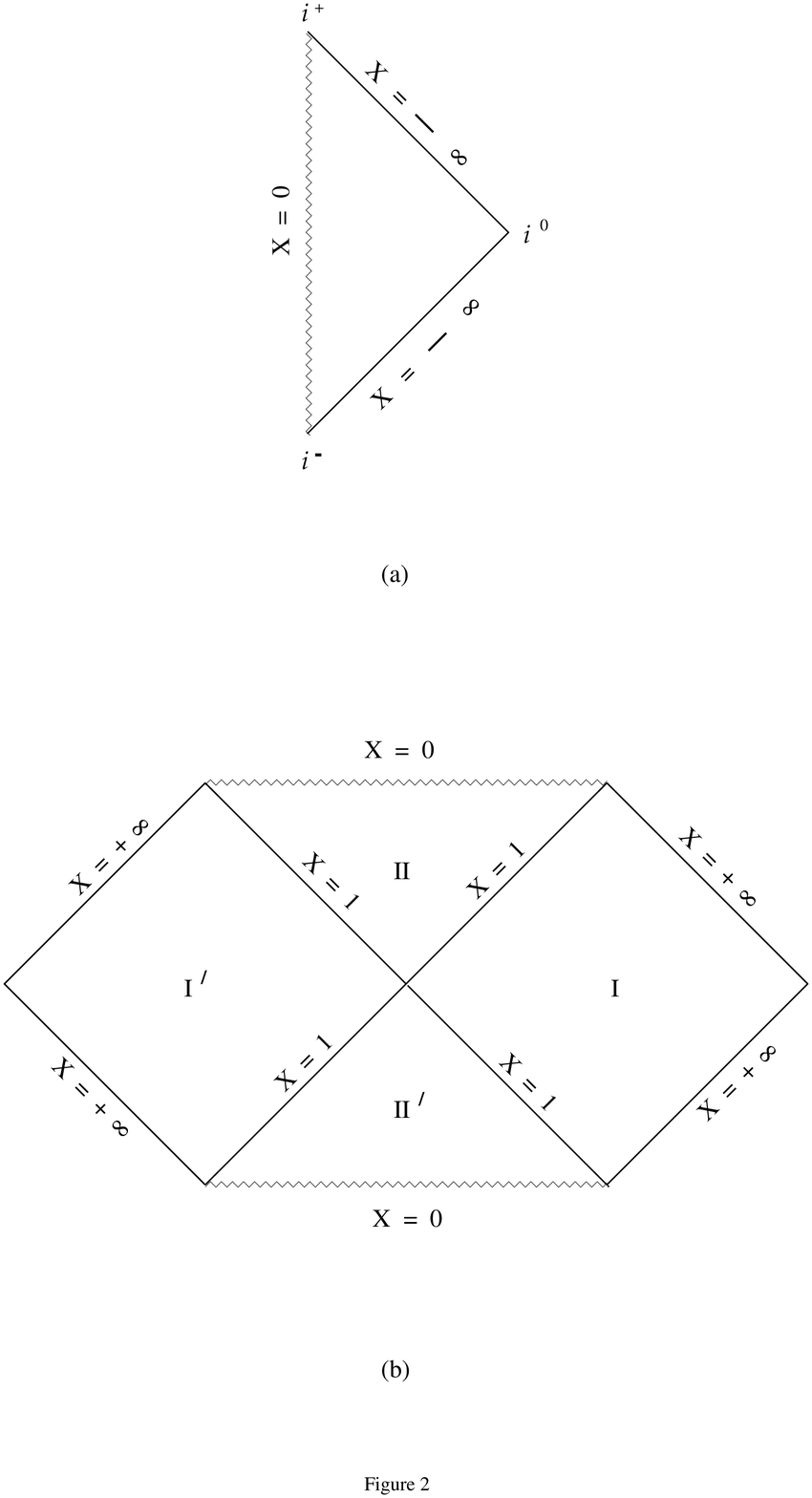}}


\end{document}